\documentclass[twoside,leqno,twocolumn]{article}
\usepackage[letterpaper]{geometry}
\usepackage{ltexpprt}
\usepackage{hyperref}

\usepackage[noend]{algpseudocode}
\usepackage[title]{appendix}
\usepackage{amsmath}
\usepackage{amssymb}
\usepackage{mathrsfs} 
\usepackage[noabbrev,capitalize,nameinlink]{cleveref}
\usepackage{color, colortbl, xcolor}
\usepackage{import}

\usepackage{float}
\usepackage{hyperref}
\usepackage{breakurl}

\usepackage{bm}
\usepackage{csquotes}
\usepackage{graphicx}
\usepackage{multirow}
\usepackage{pgfplots}
\usepackage{soul}
\usepackage{tikz}

\graphicspath{{./graphics/}}
\pgfplotsset{compat=1.18}

\definecolor{tableHeader1}{HTML}{656565}
\definecolor{tableHeader2}{HTML}{9F9F9F}
\definecolor{tableHeader3}{HTML}{D9D9D9}
\newcolumntype{C}[1]{>{\centering\let\newline\\\arraybackslash\hspace{0pt}}m{#1}}

\creflabelformat{equation}{#2\textup{#1}#3}
\Crefname{equation}{Equation}{Equations}
\Crefname{table}{Table}{Tables}
\Crefname{figure}{Figure}{Figures}
\Crefname{@algorithm}{Algorithm}{Algorithms}
\Crefname{ALC@unique}{Line}{Lines}

\DeclareSymbolFontAlphabet{\amsmathbb}{AMSb}%
\newcommand{\Exp}{\amsmathbb{E}}
\newcommand{\Var}{\mathrm{Var}}
\newcommand{\bigo}{\mathcal{O}}

\usetikzlibrary{shapes.callouts}

\newcommand{\parens}[1]{\mathopen{}\left( {#1}_{{}_{}}\,\negthickspace\right)\mathclose{}}

\newcommand{\lcsLen}[2]{\operatorname{\mathcal{L}}(#1, #2)}
\newcommand{\head}[1]{h(#1)}
\newcommand{\tail}[1]{T(#1)}

\newcommand{\define}[1]{\textit{\textbf{#1}}}
\newcommand{\codevar}[1]{\textup{\texttt{#1}}}

\creflabelformat{equation}{#2\textup{#1}#3}
\Crefname{equation}{Equation}{Equations}
\Crefname{table}{Table}{Tables}
\Crefname{figure}{Figure}{Figures}

\begin{document}

\newcommand\relatedversion{}

\title{\Large Improved Lower Bounds on the Expected Length of Longest Common Subsequences\relatedversion}
\author{
    George T. Heineman\thanks{Worcester Polytechnic Institute, United States} \and
    Chase Miller \and
    Daniel Reichman \and
    Andrew Salls \and
    G\'{a}bor S\'{a}rk\"{o}zy \and
    Duncan Soiffer
}
\date{}
\maketitle






\begin{abstract}
    It has been proven \cite{ChvatalSankoff1975} that, when normalized by $n$, the expected length of a longest common subsequence of $d$ random strings of length $n$ over an alphabet of size $\sigma$ converges to some constant that depends only on $d$ and $\sigma$. These values are known as the Chv\'{a}tal-Sankoff constants, and determining their exact values is a well-known open problem. Upper and lower bounds are known for some combinations of $\sigma$ and $d$, with the best lower and upper bounds for the most studied case, $\sigma=2, d=2$, at $0.788071$ and $0.826280$, respectively \cite{Lueker2009}. Building off previous algorithms for lower-bounding the constants, we implement runtime optimizations, parallelization, and an efficient memory reading and writing scheme to obtain an improved lower bound of $0.792665992$ for $\sigma=2, d=2$. We additionally improve upon almost all previously reported lower bounds for the Chv\'{a}tal-Sankoff constants when either the size of alphabet, the number of strings, or both are larger than 2.
\end{abstract}

\section{Introduction} \label{sec:intro}
A \define{subsequence} of a given string is a string obtained by removing zero or more characters
. Given two strings $a, b$ over an alphabet $\Sigma$, a \define{Longest Common Subsequence} (LCS) of the two strings is the longest string that occurs as a subsequence of both $a$ and $b$. Computing an LCS is a fundamental algorithmic problem in computer science and is solvable using a classical text book example of a polynomial-time dynamic programming algorithm (see, e.g., \cite{Bellman2010, Dasgupta2006}).
The LCS problem is relevant to many different fields including computational biology \cite{Huo2021, SankoffCedergren1973}, cryptography \cite{Changder2010, RoyChangder2014}, and computational linguistics \cite{Hunt2008, Levenshtein1965, MasekPaterson1980}, and sees applications in remarkably diverse areas including data comparison tools, revision control systems, stratigraphy, and even in the mathematical analysis of bird songs \cite{Hunt2008, KruskalSankoff1983}.

A major open problem related to longest common subsequences concerns determining the expected length of an LCS of $d$ random strings of length $n$ whose characters are chosen independently and uniformly over an alphabet of size $\sigma$. It has been shown that as $n$ goes towards infinity, the expected LCS length normalized by $n$ converges to a constant that depends only on $\sigma$ and $d$ \cite{ChvatalSankoff1975, KiwiSoto2009}.
These constants are called the Chv\'{a}tal-Sankoff constants and their exact values are unknown \cite{ChvatalSankoff1975}. The Chv\'{a}tal-Sankoff constants were originally defined for an LCS of only two strings, but were later expanded for an arbitrary number of strings $d$. We denote the generalized constants as $\gamma_{\sigma, d}$, and use $\gamma$ or write simply `the Chv\'{a}tal-Sankoff constant' when referring to the constant for two binary strings. Formally,
\begin{Definition}
    The \define{Chv\'{a}tal-Sankoff constant $\gamma_{\sigma, d}$} is defined as
    \begin{equation*}
        \gamma_{\sigma, d} = \lim\limits_{n \to \infty} \frac{\Exp(X_{\sigma, d, n})}{n}
    \end{equation*}
    where $X_{\sigma, d, n}$ is a discrete random variable for the length of a longest common subsequence of $d$ strings of length $n$ whose characters are independently and uniformly selected from an alphabet with $\sigma$ symbols.
\end{Definition}

The problem of calculating the Chv\'{a}tal-Sankoff constants has received significant attention and is explicitly mentioned in several textbooks (see \cite{Pevzner2000, Steele1997, SzpankowskiWojciech2001, Waterman1995}). Accordingly, extensive effort has been put into approximating and calculating lower and upper bounds for $\gamma$ and for various $\gamma_{\sigma, d}$ (see \cref{sec:background-related-work}). In continuing these efforts, we improve upon nearly all previously reported lower bounds on the Chv\'{a}tal-Sankoff constants, with particular attention towards $\gamma_{2,2}$, bringing its lower bound from $0.788071$ up to $0.792665992$.

\noindent \textbf{Our Approach.} To improve on previous lower bounds, we build off of algorithms for generating lower bounds by Lueker \cite{Lueker2009} and Kiwi and Soto \cite{KiwiSoto2009}. We take advantage of the independent nature of the performed calculations to parallelize the algorithms, and implement them from scratch in C++ with several runtime optimizations. We focus primarily on $\sigma=2, d=2$, where our changes speed up the algorithm substantially and allow us to compute values for larger string lengths than previously possible. In addition to parallelization, we employ an alternate method of encoding pairs of binary strings that allows us to pre-compute a portion of each iteration. Using this encoding, we also develop an efficient recursive strategy for sequentially reading and writing to external memory, considerably reducing the memory I/O bottleneck. Additionally, by combining several parts of the algorithm at the implementation level, we are able to reuse computations and significantly lessen how many memory I/O operations are required. In combination with advancements in compute power since the publication of previous algorithms, these improvements allowed us to achieve new state-of-the-art lower bounds for, to the best of our knowledge, all but $\gamma_{14,2}$.


\section{Background and Related Work} \label{sec:background-related-work}
In 1975, Chv\'{a}tal and Sankoff \cite{ChvatalSankoff1975} began investigating the expected length of a longest common subsequence of two strings of length $n$ whose characters are selected independently and uniformly at random from an alphabet of size $k$. They observed that, for fixed $k$, the expected length of a longest common subsequence of the strings is superadditive with respect to $n$. Using this, they proved that the expected LCS length of the strings, normalized by $n$, converges to some constant $c_k$ as $n$ grows to infinity. This constant is now referred to as the Chv\'{a}tal-Sankoff constant $\gamma_{k, 2}$. The exact values for these constants are still unknown, but continuous progress upon upper and lower bounds has been made (see \cref{tab:bounds-estimates}).

\begin{table*}[ht]
    \centering
    \caption{History of bounds and estimates of the Chv\'{a}tal-Sankoff constant, $\gamma$.}
    \begin{tabular}{|c|c|c|c|} \hline 
         \textbf{Researchers} & \textbf{Year} & \textbf{Proven Bounds} & \textbf{Estimates} \\ \hline 
         Chv\'{a}tal and Sankoff \cite{ChvatalSankoff1975} & 1975 & $0.697844 \leq \gamma \leq 0.866595$ & $\gamma \approx 0.8082$ \\
         Deken \cite{Deken1979} & 1979 & $0.7615 \leq \gamma \leq 0.8575$ & \\
         Steele \cite{Steele1986} & 1986 & & $\gamma \overset{?}{=} \frac{2}{1+\sqrt{2}} \approx 0.8284$ \\
         Dan\v{c}\'{i}k and Paterson \cite{Dancik1994, DancikPaterson1995} & 1995 & $0.77391 \leq \gamma \leq 0.83763$ & 
         \\
         Boutet de Monvel \cite{BoutetdeMonvel1999} & 1999 & & $\gamma \approx 0.812282$ \\
         Baeza-Yates et al. \cite{Baeza-Yates1999} & 1999 & & $\gamma \approx 0.8118$\\
         Bundschuh \cite{Bundschuh2001} & 2001 & & $\gamma \approx 0.812653$ \\
         Lueker \cite{Lueker2009} & 2009 & $0.788071 \leq \gamma \leq 0.82628$ & \\
         Bukh and Cox \cite{BukhCox2022} & 2022 & & $\gamma \approx 0.8122$ \\ \hline
    \end{tabular}
    \label{tab:bounds-estimates}
\end{table*}

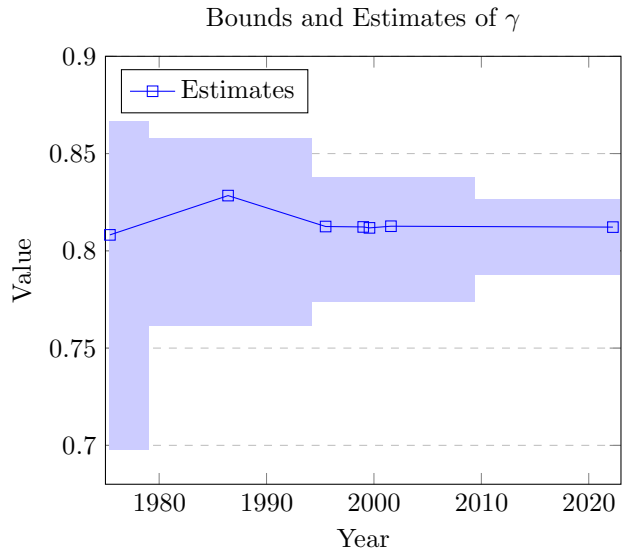
\begin{figure}[ht]
    \centering
    \begin{tikzpicture}
        \begin{axis}[
            title={Bounds and Estimates of $\gamma$},
            xlabel={Year},
            ylabel={Value},
            xmin=1975, xmax=2023,
            ymin=0.68, ymax=0.9,
            legend pos=north west,
            ymajorgrids=true,
            grid style=dashed,
            /pgf/number format/.cd,
            1000 sep={}
        ]
        
        \shade[left color=blue!20, right color=blue!20] (1975.4167,0.697844) rectangle (1979,0.866595);
        \shade[left color=blue!20, right color=blue!20] (1979,0.7615) rectangle (1994.166,0.8575);
        \shade[left color=blue!20, right color=blue!20] (1994.166,0.77391) rectangle (2009.3333,0.83763);
        \shade[left color=blue!20, right color=blue!20] (2009.3333,0.788071) rectangle (2022.9,0.82628);

        \addplot[
            color=blue,
            mark=square,
            ]
            coordinates {
            (1975.4167, 0.8082)(1986.4167, 0.8284)(1995.5, 0.8125)(1999, 0.812282)(1999.5833, 0.8118)(2001.5833, 0.812653)(2022.25, 0.8122)
            };
            \legend{Estimates}
        \end{axis}
    \end{tikzpicture}
    \caption{The best upper and lower bounds on $\gamma$ and estimates of $\gamma$ over time.}
\end{figure}
Chv\'{a}tal and Sankoff established in their original paper that the Chv\'{a}tal-Sankoff constant for the binary case (i.e., $\gamma$) falls between $0.697844$ and $0.866595$ \cite{ChvatalSankoff1975}. Through Monte-Carlo simulations, they predicted the value to be around $0.8082$. Deken \cite{Deken1979} initiated efforts to refine these bounds in 1979, culminating in 1983 with a tighter interval of $0.7615$ to $0.8575$, achieved via diverse matching algorithms. Despite their differences, Deken's algorithms adhere to a common procedural framework: they initialize a pointer on each sequence, alternate movements until encountering an increase in the count of differing elements between the pointers, match characters when possible, and resume the process from the matched points, disregarding any prior characters.

In 1986, Steele \cite{Steele1986} conjectured that the exact value of the binary Chv\'{a}tal-Sankoff constant was $\frac{2}{1+\sqrt{2}}$ ($\approx 0.82843$). This conjecture stemmed from a heuristic approach inspired by coin-flip sequences, assuming an independent and identical distribution of each coin flip. Much of this conjecture's development was spurred by the contributions of Arratia and Waterman \cite{ArratiaWaterman1985a, ArratiaWaterman1985b}.

Dan\v{c}\'{i}k and Paterson \cite{DancikPaterson1995} introduced a novel method in 1994. They defined a system of linear recurrence relations, where each relation counts the number of string pairs of length $m$ with an LCS of length $i$. Each linear recurrence additionally requires that the string pairs start with a specific pattern. They then showed that the maximum eigenvalue of the system of recurrence relations can be used to determine an upper bound $\gamma$. By minimizing the overlap between different recurrence relations, the upper bound on $\gamma$ is improved. In their best results, Dan\v{c}\'{i}k and Paterson used 52 linear recurrences, obtaining a best upper bound of $0.837623$. Separately, they also calculated a lower bound of $0.77391$ using analysis on finite state machines that generate the length of an LCS, given two strings as inputs \cite{Dancik1994}.

Boutet de Monvel \cite{BoutetdeMonvel1999}, in early 1999, conducted extensive Monte-Carlo simulations akin to Chv\'{a}tal and Sankoff, albeit optimized, parallelized, and with increased sample size. His prediction for the binary constant stood at 0.812282. Later that year, Baeza-Yates, Gavald\`{a}, Navarro, and Scheihing \cite{Baeza-Yates1999} estimated the constant at 0.8118, leveraging finite state automata and analysis of string generation complexity (via Kolmogorov complexity analysis). In 2001, Bundschuh \cite{Bundschuh2001} employed a lattice framework to tackle the longest common subsequence problem, resulting in an approximated value of 0.812653.

In 2009, Lueker \cite{Lueker2009} utilized dynamic programming to create a modified version of Dan\v{c}\'{i}k and Paterson's algorithm which performed calculations on every pair of strings of length $\ell$, to refine the bounds on $\gamma$ to between 0.788071 and 0.82628. This disproved Steele's conjectured value for the binary constant, as Lueker's upper bound fell below Steele's prediction. That same year, Kiwi and Soto \cite{KiwiSoto2009} generalized Lueker's lower-bound algorithm to compute lower bounds for a wider range of alphabets and numbers of strings. Along with some results obtained for $\gamma_{\sigma, 2}$ for various values of $\sigma >2$ (see \cite{Dancik1994, Dancik1998, DancikPaterson1995, Deken1979}), these were the state-of-the-art results before our project.

In 2022, Bukh and Cox \cite{BukhCox2022} used a variant of PushTASEP
to obtain results related to the LCS problem.
They first define $\Delta(2n) = \lcsLen{st}{uv} - \lcsLen{s}{u} - \lcsLen{t}{v}$ where $s,t,u,v$ are independent, random strings of length $n$ and $\lcsLen{x}{y}$ denotes the length of a longest common subsequence of the strings $x$ and $y$. They propose a conjecture asserting the existence of constants $c_{1}$ and $c_{2}$ such that $\Exp(\Delta(n)) \sim c_{1}\sqrt[3]{n}$ and $\sqrt{\Var(\Delta(n))} \sim c_{2}\sqrt[3]{n}$. Bukh and Cox also performed Monte-Carlo simulations with a probabilistic $\bigo({n^{\frac{3}{2}}})$ LCS approximation heuristic, which they used to conjecture constants $c_{1} \approx \frac{1}{2}$ and $c_{2} \approx \frac{1}{4}$ \cite{BukhCox2022}. Using their conjecture, they estimate the binary Chv\'{a}tal-Sankoff constant as 0.8122.

Also of note is work by  Kiwi, Loebl, and Matoušek \cite{Kiwi2005} in 2005 which proves, via subadditivity arguments, that $\lim\limits_{\sigma \to \infty} \gamma_{\sigma,2}\sqrt{\sigma}= 2$. Additionally, in 2022, Tiskin \cite{Tiskin2022} demonstrated that, by linking $\gamma$ to the parameters of a specific stochastic particle process, $\gamma$ is the solution to an explicit (but intractably large) system of polynomial equations with integer coefficients.

\subsection{The Kiwi-Soto Algorithm}

Here we introduce the algorithm of Kiwi and Soto \cite{KiwiSoto2009} for calculating lower bounds on the Chv\'{a}tal-Sankoff constants, which we used with several modifications to obtain our results. This algorithm is based on the overall process described by Lueker \cite{Lueker2009}, but generalized for arbitrary values of $\sigma$ and $d$. We begin by introducing some notation and summarizing the theory behind the algorithm.

Throughout this paper, we adopt the convention that all vectors are represented in bold and that a numerical value in bold is a vector where every coordinate is that value. Additionally, all vectors contain real numbers and are of length $\sigma^{d \cdot \ell}$, corresponding to the number of possible $d$-tuple of strings where each string has $\ell$ characters taken from an alphabet of size $\sigma$. Vector inequalities are evaluated coordinate-wise (e.g., we say $\mathbf{v_1} \leq \mathbf{v_2}$ if $\mathbf{v_1}[i] \leq \mathbf{v_2}[i]$ for all $i$). We also introduce two new definitions.
\begin{Definition}
    For a string $a$, we define the first character of $a$ to be the \define{head} of $a$, denoted as $h(a)$. We define the rest of $a$ to be the \define{tail} of $a$, denoted $T(a)$. Note that $a = h(a)T(a)$.
\end{Definition}

Following Kiwi and Soto's notation, let $x_1, \ldots, x_d$ be a collection of $d$ random strings of length $n$ chosen independently and uniformly from an alphabet $\Sigma$ of size $\sigma$, let $a_1,\ldots,a_d$ be a collection of $d$ finite strings over the same alphabet, let $L$ be the function that returns the length of the LCS of two or more strings, and for any string $s$ let $s[..i]$ denote the substring formed by the first $i$ characters of $s$. Finally, let \[ W_n(a_1,\ldots,a_d) = \Exp\left[\max\limits_{i_1 + \cdots + i_d =n} L(a_1x_1[..i_1],\ldots,a_dx_d[..i_d])\right]\]
This quantity is the expected length of an LCS of $d$ strings with prefixes $a_1, \ldots, a_d$, respectively, and $d$ suffixes whose lengths sum to $n$ and whose characters are independently and uniformly chosen from $\Sigma$ \cite{KiwiSoto2009}. Kiwi and Soto show that, for all $a_1,\ldots,a_d$,
\begin{equation} \label{eq:limequals}
    \gamma_{\sigma,d} = \lim_{n\to\infty} \frac{W_{nd}(a_1,\ldots,a_d)}{n}.
\end{equation}
Since this quantity is increasing, by lower-bounding it we can also lower-bound $\gamma_{\sigma,d}$. This is now our goal.

Fix $\ell$ in $\mathbb{N}$ and let $\mathbf{w_n}$ denote the $\sigma^{\ell d}$-dimensional vector whose coordinates are the values of $W_{nd}(a_1,\ldots,a_d)$ when $a_1,\ldots,a_d$ each vary over all strings in $\Sigma^\ell$ (the $\ell$-ary Cartesian power of $\Sigma$). Kiwi and Soto establish a lower bound for each coordinate of $\mathbf{w_n}$ as a function of vectors $\mathbf{w_m}$, with $m < n$.

To start, if all the strings $a_1,\ldots,a_d$ begin with the same character, then the following inequality holds:
\begin{equation}\label{eq:ineq1}
    \begin{split}
    & \mathbf{w_n}[a_1,\ldots,a_d] \geq \\ & 1 + \frac{1}{|\Sigma^d|}\sum_{(c_1,\ldots,c_d) \in \Sigma^d}\mathbf{w_{n-d}}[T(a_1)c_1,\ldots,T(a_d)c_d]
    \end{split}
\end{equation}
This inequality states that if all the strings start with the same character, then the expected length of the LCS of all of them, permitting $n$ random extra characters, is at least 1 (the first character) plus the average of the expected length of the LCS of the strings obtained by removing the first character and adding $d$ of the $n$ random characters \cite{KiwiSoto2009}.

If all the strings do \textit{not} begin with the same character, then we must take a more complicated approach.
For this purpose, we define a function $F_z$ which computes the expected length of an LCS of the strings $A=(a_1, \dots, a_d)$ when discarding the first character of all strings not beginning with $z$ and adding to their end a random character from $\Sigma$.
\begin{algorithm}{Calculating $F_z$}
    \label{alg:fz}
    \hfill
    \begin{algorithmic}[1]
        \Function{$F_z$}{$\mathbf{v_1}, \dots, \mathbf{v_d}, A$}
            \State $r \gets 0$
            \State $N \gets ( j\: |\: j \in (1, \dots, d), \; h(s_j) \neq z )$
            \For{$(c_1,\ldots,c_{|N|}) \in \Sigma^{|N|}$}
                \State $(s_1,\ldots,s_d) \gets (a_1,\ldots,a_d)$
                \For{$j \in \{1, \ldots, |N|\}$}
                    \State $s_{N[j]} \gets T(s_{N[j]})c_j$
                \EndFor
                \State $r \gets r + \mathbf{v_{|N|}}[s_1, \dots, s_d]$ 
            \EndFor
            \State \Return $\sigma^{-|N|} \cdot r$
        \EndFunction
    \end{algorithmic}
\end{algorithm}

\noindent Here, $N$ is an ordered set containing the positions of strings in the tuple $(a_1, \dots, a_d)$ that do \textit{not} start with some character $z$. $F_z$ removes the first character from all strings that do not start with $z$ and adds some new character to their ends, doing this for all possible combinations of new characters at the end of the strings. $F_z$ then calculates the mean value of $\mathbf{v_{|N|}}$ at the indices corresponding to all $\sigma^{|N|}$ of the $d$-tuples created this way.

Using this, for all $d$-tuples of strings $A \in (\Sigma ^ \ell)^d$, we have that 
\begin{equation}\label{eq:ineq2}
    \mathbf{w_n}[A] \geq \max\limits_{z \in \Sigma} F_z(\mathbf{w_{n-d}}, \dots, \mathbf{w_{n-1}}, A).
\end{equation}

If we let $b$ have value $1$ if all strings in $A$ start with the same character and have value $0$ otherwise, we can combine inequalities (\ref{eq:ineq1}) and (\ref{eq:ineq2}) into a single equation:
\begin{equation}\label{eq:ineq3}
    \mathbf{w_n}[A] \geq b + \max\limits_{z \in \Sigma} F_z(\mathbf{w_{n-1}}, \dots, \mathbf{w_{n-d}}, A).
\end{equation}

For notational convenience, let $F \colon (\mathbb{R}^{\sigma^{d \ell}})^d \rightarrow \mathbb{R}^{\sigma^{d \ell}}$ be the function that takes as input a collection of vectors $\mathbf{v_{n-1}}, \dots, \mathbf{v_{n-d}}$ and outputs the vector whose coordinates are given by $b + \max\limits_{z \in \Sigma} F_z(\mathbf{v_{n-1}}, \dots, \mathbf{v_{n-d}}, A)$ as $A$ varies over all string tuples in $(\Sigma^\ell)^d$.
To turn (\ref{eq:ineq3}) into a useful lower bound on $\gamma_{\sigma,d}$, Kiwi and Soto prove the following pivotal lemma:
\begin{lemma}
\label{lem:fdef}
Suppose a function $F: (\mathbb{R}^{\sigma^{d  \ell}})^d \mapsto \mathbb{R}^{\sigma^{d \ell}}$ satisfies the following three properties:
    \begin{enumerate}
        \item {\normalfont\textbf{Monotonicity.}} Given two d-tuples of vectors $(\mathbf{v_1}, \dots \mathbf{v_d}) \leq (\mathbf{u_1}, \dots, \mathbf{u_d})$ component-wise, then $F(\mathbf{v_1}, \dots \mathbf{v_d}) \leq F(\mathbf{u_1}, \dots, \mathbf{u_d})$.
        
        \item {\normalfont\textbf{Translation Invariance.}} For all $r \in \mathbb{R}$, $F(\mathbf{v_1} + r\mathbf{1}, \dots, \mathbf{v_d} + r\mathbf{1}) = F(\mathbf{v_1}, \dots, \mathbf{v_d}) + r\mathbf{1}$.
        
        \item {\normalfont\textbf{Feasibility.}} There exists a \define{feasible triplet} for F. That is, there exists a 3-tuple $(\mathbf{u} \in \mathbb{R}^{\sigma^{d \ell}}, r \in \mathbb{R}, \epsilon \in [0, r])$ such that
        \[F(\mathbf{u} + (d - 1)r\mathbf{1}, \mathbf{u} + (d - 2)r\mathbf{1}, \dots, \mathbf{u}) \geq \mathbf{u} + (dr - \epsilon)\mathbf{1}.\]
    \end{enumerate}
    Then, for any sequence of $(\mathbf{v_n})_{n\in \mathbb{N}}$ of vectors in $\mathbb{R}^{\sigma ^{d \ell}}$ such that $\mathbf{v_n} \geq F(\mathbf{v_{n-1}}, \ldots, \mathbf{v_{n-d}})$ for all $n \geq d$, there exists a vector $\mathbf{u_0} \in \mathbb{R}^{\sigma ^{d \ell}}$ such that for all $n \geq 0$,
    \begin{equation}
        \mathbf{v_n} \geq \mathbf{u_0} + n(r - \epsilon)\mathbf{1}.
    \end{equation}
\end{lemma}

It follows easily from $F$'s definition that $F$ is monotone and translation invariant. Thus, if we find a feasible triplet $(\mathbf{u}, r, \epsilon)$ for $F$, then the sequence of vectors $(\mathbf{w_n})_{n\in \mathbb{N}}$ must satisfy $\mathbf{w_n} \geq \mathbf{u_0} + n(r-\epsilon)\mathbf{1}$ for all $n$ \cite{KiwiSoto2009}. From (\ref{eq:limequals}) it follows that
\begin{equation}\label{eq:lwrbound}
    \gamma_{\sigma,d} \geq d(r-\epsilon).
\end{equation}

Accordingly, to establish a good lower bound we need only find a feasible triplet where $(r-\epsilon)$ is as large as possible.
Empirically, Kiwi and Soto observed that for any set of initial vectors $\mathbf{v_1}, \ldots, \mathbf{v_{d}}$, if $\mathbf{v_{n+d}}$ is set to $F(\mathbf{v_{n+d-1}}, \ldots, \mathbf{v_n})$ for all $n \in \mathbb{N}$, then there exists a real value $r$ such that $\mathbf{v_{n+1}}-\mathbf{v_n} \approx r\mathbf{1}$ for all $n$ large enough. So, by definition of $\mathbf{v_{n+d}}$,
\begin{equation*}
    F(\mathbf{v_n} + (d-1)r\mathbf{1}, \mathbf{v_n} + (d-2)r\mathbf{1}, \ldots, \mathbf{v_n}) \approx \mathbf{v_n} + dr\mathbf{1}.
\end{equation*}
It follows that one way to find a feasible triplet for $F$ is to find $n$ large enough such that $\mathbf{v_{n+1}}-\mathbf{v_{n}} \approx x\mathbf{1}$ for some real value $x$. Then, set $\mathbf{u} = \mathbf{v_{n+1}}$ and define $r$ as the maximum value where $\mathbf{v_{n+1}} - \mathbf{v_{n}} \geq r\mathbf{1}$ holds and $\epsilon$ as the minimum value such that $(\mathbf{u}, r, \epsilon)$ is a feasible triplet for $F$ \cite{KiwiSoto2009}. This is exactly the approach taken by the Feasible Triplet Algorithm (\cref{alg:feasibletriplet}).

\begin{algorithm}{The Feasible Triplet Algorithm}
    \label{alg:feasibletriplet}
    \hfill
    \begin{algorithmic}[1]
        \Function{FeasibleTriplet}{$\sigma$, $d$, $\ell$, $n$}
            \State  $\mathbf{v} \gets \mathbf{0}$ \Comment{$\mathbf{0}$ is a $(d+1)\times\sigma^{d \ell}$ matrix of zeros}
            \State  $(\mathbf{u}, r, \epsilon) \gets (\mathbf{v_0}, 0, 0)$ \Comment{$\mathbf{v_0}$ is the 0th row of $\mathbf{v}$}
            \For{$i = d, \dots, n$}
                \State $\mathbf{v_d} \gets F(\mathbf{v_{d-1}}, \mathbf{v_{d - 2}}, \dots, \mathbf{v_0})$
                \State $R \gets \max\limits_{0 \leq j < \sigma^{d \ell}} (\mathbf{v_d} - \mathbf{v_{d-1}})[j]\footnote{$[j]$ denotes the $j$th coordinate of the vector}$
                \State $\mathbf{W} \gets \mathbf{v_d} + dR \mathbf{1} - F(\mathbf{v_d} + (d - 1) R \mathbf{1}, \mathbf{v_d} + (d - 2) R \mathbf{1}, \dots, \mathbf{v_d})$
                \State $E \gets \max \{0, \max\limits_{0 \leq j < \sigma^{d  \cdot  \ell}} \mathbf{W}[j]\}$\;
                \If{$R - E \geq r - \epsilon$}
                    \State $(\mathbf{u}, r, \epsilon) \gets (\mathbf{v_d}, R, E)$
                \EndIf
                \State \textup{Shift the rows of $\mathbf{v}$ up by 1}
            \EndFor
            \State \Return $(\mathbf{u}, r, \epsilon)$\;
        \EndFunction
    \end{algorithmic}
\end{algorithm}
Once \cref{alg:feasibletriplet} has terminated, values $r$ and $\epsilon$ can be used to calculate a lower bound on $\gamma_{\sigma, d}$ according to (\ref{eq:lwrbound}).

\subsubsection{The Binary Case}
The algorithm and related definitions can be greatly simplified (see \cref{alg:bft}) when working only with pairs of strings with an alphabet of size two (that is, $d=2, \sigma=2$). Let $a$ and $b$ be strings of length $\ell$, and let $s$ have value 1 if $a$ and $b$ start with the same character and have value 0 otherwise. If we let $F_0$ and $F_1$ be sub-functions that simplify $F_z$ to the binary case, then, at coordinate $(a, b)$, $F$ reduces to 
\begin{equation}
    \begin{split}
    &F(\mathbf{v_1}, \mathbf{v_2})[a, b] = \\ & \mathbf{s} + \max\parens{F_0(\mathbf{v_1}, \mathbf{v_2}, a, b), F_1(\mathbf{v_1}, \mathbf{v_2}, a, b)}.
    \end{split}
\end{equation}

\begin{algorithm}{The Binary Feasible Triplet Algorithm}
    \label{alg:bft}
    \hfill
    \begin{algorithmic}[1]
        \Function{BinaryFeasibleTriplet}{$\ell$, $n$}
            \State $\mathbf{v_0} \gets \mathbf{0}$ \Comment{$\mathbf{0}$ is a vector containing $2^{2 \ell}$ zeroes}
            \State $\mathbf{v_1} \gets \mathbf{0}$
            \State $(\mathbf{u}, r, \epsilon) \gets (\mathbf{v_0}, 0, 0)$
            \For{$i = 2, \dots, n$}
                \State $\mathbf{v_2} \gets F(\mathbf{v_1}, \mathbf{v_0})$
                \State $R \gets \max\limits_{0 \leq j < 2^{2 \cdot \ell}} (\mathbf{v_2} - \mathbf{v_1})[j]$
                \State $\mathbf{W} \gets \mathbf{v_2} + 2R\mathbf{1} - F(\mathbf{v_2} + R\mathbf{1}, \mathbf{v_2})$
                \State $E \gets \max\{0, \max\limits_{0 \leq j < 2^{2 \ell}} \mathbf{W}[j]\}$
                \If{$R - E \geq r - \epsilon$}
                    \State $(\mathbf{u}, r, \epsilon) \gets (\mathbf{v_2}, R, E)$
                \EndIf
                \State $\mathbf{v_0} \gets \mathbf{v_1}$\;
                \State $\mathbf{v_1} \gets \mathbf{v_2}$\;
            \EndFor
            \State \Return $(\mathbf{u}, r, \epsilon)$
        \EndFunction
    \end{algorithmic}
\end{algorithm}

We now define $F_1$ and $F_0$. Note that $F_0$ has the same definition as $F_1$, but swaps 0 and 1 when evaluating $\head{a}$ and $\head{b}$:
\begin{equation}\label{relwork:F1}
    \begin{aligned}
     & F_1(\mathbf{v_1}, \mathbf{v_2}, a, b) = &\\ & \left\{\begin{array}{lrl}
        0, & \head{a} = 1 & \head{b} = 1 \\
        
        \frac{1}{2} (\mathbf{v_1}[a, \tail{b}0] + \mathbf{v_1}[a, \tail{b}1]) & \head{a} = 1 & \head{b} = 0 \\
        
        \frac{1}{2} (\mathbf{v_1}[\tail{a}0, b] + \mathbf{v_1}[\tail{a}1, b]) & \head{a} = 0 & \head{b} = 1 \\

        \frac{1}{4} (\sum\limits_{c_1, c_2 \in \{0, 1\}} \mathbf{v_2}[\tail{a}c_1, \tail{b}c_2]) & \head{a} = 0 & \head{b} = 0
        \end{array}\right.
    \end{aligned}
\end{equation}

\begin{equation}\label{relwork:F0}
    \begin{aligned}
     & F_0(\mathbf{v_1}, \mathbf{v_2}, a, b) = &\\ & \left\{\begin{array}{lrl}
        0, & \head{a} = 0 & \head{b} = 0 \\
        
        \frac{1}{2} (\mathbf{v_1}[a, \tail{b}0] + \mathbf{v_1}[a, \tail{b}1]) & \head{a} = 0 & \head{b} = 1 \\
        
        \frac{1}{2} (\mathbf{v_1}[\tail{a}0, b] + \mathbf{v_1}[\tail{a}1, b]) & \head{a} = 1 & \head{b} = 0 \\

        \frac{1}{4} (\sum\limits_{c_1, c_2 \in \{0, 1\}} \mathbf{v_2}[\tail{a}c_1, \tail{b}c_2]) & \head{a} = 1 & \head{b} = 1
        \end{array}\right.
    \end{aligned}
\end{equation}
Essentially, when one or more of the strings starts with 0, $F_1$ returns the average of all permutations of those strings where we remove the first character and append some new character (0 or 1). When all strings start with 1, the average is 0, since there are no strings to permute. $F_0$ does the same but for strings that start with a 1.
\section{Implementation Details} \label{sec:implem}
Since a value must be computed for every possible $d$-tuple of strings over an alphabet of size $\sigma$, the Feasible Triplet Algorithm (\cref{alg:feasibletriplet}) has a time complexity of $\mathcal{O}(n\sigma^{d\ell})$, and since it must store $d+1$ vectors of size $\sigma^{d\ell}$, it has a space complexity of $\mathcal{O}(d\sigma^{d\ell})$. As the values of $\sigma$, $d$, and $\ell$ increase, this exponential scaling quickly becomes a barrier to progress. In this section, we describe our approach to overcome some of these limitations. We focus primarily on optimizing the Binary Feasible Triplet Algorithm; only parallelization is implemented as an improvement to both the general and binary versions of the Feasible Triplet Algorithm.

\subsection{Parallelization}
It can be observed that in \cref{alg:feasibletriplet}, each coordinate of $\mathbf{v_d}$ can be computed independently of every other coordinate of $\mathbf{v_d}$. That is, for all string tuples $A \in (\Sigma^\ell)^d$, $\mathbf{v_d}[A]$ can be computed using only vectors $\mathbf{v_{d-1}}, \ldots, \mathbf{v_{0}}$. This same reasoning applies to the computation of $\mathbf{W}$. Accordingly, the algorithm is amenable to parallelization. To implement this, a chosen number of threads is spun up, with each thread calculating the values of $\mathbf{v_d}$ for a distinct slice of the vector. This optimization is applied to both the binary and general versions of the Feasible Triplet Algorithm; the remainder of this section concerns only the algorithm for $\sigma=2, d=2$. Material in \cref{appx1} additionally describes how this independence can be relaxed slightly in order to reuse computations without sacrificing the ability to parallelize.

\subsection{Indexing}\label{impl:indexing}
A pair of binary strings can be represented as a single 64-bit unsigned integer so long as their combined length does not exceed 64 characters and a consistent indexing scheme is identified. Given two strings $a$ and $b$, we choose to index them by interleaving their bits starting with $a$ and filling the remaining bits with zeros on the left. For instance, a pair of strings $a = 1011$ and $b=0010$ is represented as $0...10001110$. We say a 64-bit integer is an \define{interleaved string pair}, or simply a \define{pair}, if it represents two binary strings interleaved in this fashion.

This scheme has several desirable properties. Firstly, calculating $\mathbf{v_2}$ can be done (in sequential order) by simply iterating through all integers from 0 to $2^{2\ell}-1$, which we denote $[0,...,2^{2\ell})$. Secondly, we need not check whether the first bit of $a$ and $b$ match: we know ahead of time that pairs $[0,...,2^{2\ell-2})$ have the same first bit (0) for $a$ and $b$, pairs $[2^{2\ell-2},...,2^{2\ell-1}+2^{2\ell-2})$ have different first bits, and pairs $[2^{2\ell-1}+2^{2\ell-2},...,2^{2\ell})$ have the same first bit (1).
Lastly, this indexing will allow us to determine a method of reading disk memory sequentially once arrays become too large to fit in RAM (\cref{impl:seqmem}).

\subsection{Array Reductions and Symmetries}
In the binary case, Lueker notes how it is not necessary to store arrays $\mathbf{v_i}$, for three consecutive integers $i$, in memory, as the recurrence can be simplified by storing only two arrays (that is, $\mathbf{v_1}, \mathbf{v_2}$) and iterating on those instead, with the bound calculation adjusted appropriately \cite{Lueker2009}. We also adopt this optimization. As Lueker also notes, one can further observe that since complementing both $a$ and $b$ does not impact the length of their longest common subsequence, we get that
\[
    \mathbf{v_i}[(\overline{a\vphantom{b}}, \overline{b})] = \mathbf{v_i}[(a, b)],
\]
where $\overline{a}$ represents the binary complement of $a$ \cite{Lueker2009}. With the indexing scheme defined in \cref{impl:indexing}, for any two indices $i$ and $j$ such that
\[
    2^{2\ell-1}>i\ge0 \text{ and } j=2^{2\ell}-1-i,
\]
we have that
\begin{equation}\label{impl:comp_eq}
    2^{2\ell}> j \ge 2^{2\ell-1} \text{ and } (a_i, b_i) = (\overline{a_j\vphantom{b}}, \overline{b_j}).
\end{equation}
Thus, we need only iterate strings pairs represented by the integers $[0,...,2^{2\ell-1})$.

\subsection{Sequential Memory Access}\label{impl:seqmem}
As $\ell$ grows, memory becomes a limiting factor far faster than computation time. For instance, at $\ell=20$ with 4 bytes for each value stored in the vector, a naive implementation requires
\[
    2^{2 \cdot 20}  \cdot  4 = 4 \; 398 \; 046 \; 511 \; 104 \text{ bytes} \approx 4.4 \text{ TB}
\] per vector. Even if symmetry is fully exploited, this requires roughly a terabyte of memory per vector. With our resources, it is infeasible to store all of these values in RAM.

Accordingly, vectors must be read from and written to external (disk) memory. However, operating exclusively from external memory is also infeasible: we experienced over a 100x slow down when reading and writing values only from disk. File I/O time dwarfed the time taken to actually perform the computations for the recurrence, largely because memory accesses were non-sequential. 

To alleviate this overhead, reading from and writing to disk must be done in large sequential blocks. As such, a recursive approach is taken to identify contiguous blocks of integers from $[0,...,2^{2\ell-1})$ whose accessed values also span a contiguous block of integers. This section outlines the approach taken and justifies why it is correct.

The function as implemented contains three primary loops: $L_{0, 0}$, $L_{0, 1}$, and $L_{1, 0}$. Each loop is valid only for a predetermined range of string pairs, but accepts a start parameter and end parameter to allow for subdivision of that range.

\subsubsection{\texorpdfstring{$\bm{L_{0,0}}$}{L}}
$L_{0, 0}$ iterates over string pairs whose first bits match and calculates the new value in the recurrence according to the last case of $F_1$ (see (\ref{relwork:F1})). For a particular $a,b$ string pair, with the interleaved indexing, this is accomplished by the following simple procedure:

\begin{algorithm}
    \label{alg:samefirstbit}
    \hfill \\
        \textbf{Input:} $x$, an interleaved $a$, $b$ string pair \\
        \textbf{Output:} The new recurrence value for index $x$
        \begin{algorithmic}[1]
            \Function{SameFirstBit}{$x$}
                \State Bitshift $x$ left by 2 \;
                \State \Return $1 + \frac{1}{4} (\mathbf{v_1}[x]+\mathbf{v_1}[x+1]+\mathbf{v_1}[x+2]+\mathbf{v_1}[x+3])$ \;
            \EndFunction
    \end{algorithmic}
\end{algorithm}

Since $L_{0, 0}$ iterates only within $[0,...,2^{2\ell-2})$, the first bit of $a$ and $b$ will always be 0. Thus, this procedure always accesses the values at $4x$, $4x+1$, $4x+2$, and $4x+3$. For $x\geq 2^{2\ell-3}$, it will access pairs $\geq 2^{2\ell-1}$, so we use (\ref{impl:comp_eq}) to transform the pairs back to their symmetric position within the vector. As a result, if $x$ is iterated in sequential order, $L_{0,0}$ first accesses values sequentially within $[0,...,2^{2\ell-1})$ and writes out values sequentially within $[0,...,2^{2\ell-3})$, and then accesses values sequentially within $(2^{2\ell-1},...,0]$ and writes out values sequentially within $(2^{2\ell-2},...,2^{2\ell-3}]$.

\subsubsection{\texorpdfstring{$\bm{L_{0,1}}$}{L} and its Recursion}
$L_{0, 1}$ iterates over string pairs whose first bits do not match. It calculates the second case ($h(a)=0, h(b)=1$) from $F_0$ (see (\ref{relwork:F0})). This can be accomplished by another simple procedure:

\begin{algorithm}
    \label{alg:diffFirstBit}
    \hfill \\
    \textbf{Input:} $x$, an interleaved $a$, $b$ string pair \\
    \textbf{Output:} The new recurrence value for index $x$
    \begin{algorithmic}[1]
        \Function{DifferentFirstBit}{$x$}
            \State $a \gets \text{ even bits of } x$
            \State $b \gets \text{ odd bits of } x$
            \State Zero out first bit of $b$
            \State Bitshift $b$ left by 2
            \State $i \gets a \mathbin{|} b$ \Comment{Bitwise OR recombines $a$ and $b$}
            \State \Return $\frac{1}{2} (\mathbf{v_1}[i]+\mathbf{v_1}[i+1])$
        \EndFunction
    \end{algorithmic}
\end{algorithm}

Since the string pairs are iterated only within $[2^{2\ell-2},...,2^{2\ell-1})$, the first bit of $a$ is always 0 and the first bit of $b$ is always 1. As such, the second bit of $b$ uniquely determines whether it is necessary to access values from $[0,...,2^{2\ell-2})$ or $[2^{2\ell-2},...,2^{2\ell-1})$ to calculate the new value for a particular string pair. Similarly, the second bit of $a$ uniquely determines whether it is necessary to access values from the first half or second half of the range determined by the second bit of $b$, for a total of four possible access ranges of size $2^{2\ell-3}$. For example, by looking at only the second bit of $a$ and $b$ for string pair $0...011010$ ($\ell=3$), we know we need only access values from $[2^{2\ell-3},...,2^{2\ell-2})$. In fact, we know that every string pair in the range $[2^{2\ell-2}+2^{2\ell-3},...,2^{2\ell-2}+2^{2\ell-3}+2^{2\ell-4})$ is constrained to accessing values from the range $[2^{2\ell-3},...,2^{2\ell-2})$. Additionally, since each string pair accesses two distinct values, it is guaranteed that every value in one of these ranges is accessed exactly once.

Within each of these four ranges, we can again subdivide into four blocks of equal size based on the next bit of $a$ and of $b$, again with the same guarantees on access. This subdivision can be performed recursively until the vector has been divided into chunks small enough to fit into RAM (based on a chosen \codevar{stop\_depth}). At this point, each chunk can be read \textit{sequentially} into RAM, computation performed for the chunk's corresponding values, and results written sequentially out to disk. This recursion is demonstrated in \cref{alg:recurse}.

\newpage 
\begin{algorithm}
    \label{alg:recurse}
    \hfill
    \begin{algorithmic}[1]
        \Function{Recurse}{\codevar{offset}, \codevar{idx\_offset}, \codevar{depth}}
            \State $\codevar{num\_strs} \gets 2^{2(\ell-\codevar{depth})-2}$
            \If{$\codevar{depth} < \codevar{stop\_depth}$}
                \State \Call{Recurse}{$\codevar{offset}$, $ \codevar{idx\_offset}$, $ \codevar{depth}+1$}
                \State \Call{Recurse}{$\codevar{offset} + \codevar{num\_strs}/4$, $\codevar{idx\_offset} + 2 \times \codevar{num\_strs}/4$, $\codevar{depth}+1$}
                \State \Call{Recurse}{$\codevar{offset} + 2 \times \codevar{num\_strs}/4$, $\codevar{idx\_offset} + \codevar{num\_strs}/4$, $\codevar{depth}+1$}
                \State \Call{Recurse}{$\codevar{offset} + 3 \times \codevar{num\_strs}/4$, $ \codevar{idx\_offset} + 3\times \codevar{num\_strs}/4$, $\codevar{depth}+1$}
            \Else
                \State load in $\mathbf{v_1}$[\codevar{idx\_offset}] through $\mathbf{v_1}$[$\codevar{idx\_offset} + 2 \times \codevar{num\_strs}$]
                \State $\codevar{start} \gets 2^{2\ell-2}+\codevar{offset}$
                \State $\codevar{end} \gets \codevar{start}+\codevar{num\_strs}$
                \State \Call{$L_{0,1}$}{\codevar{start}, \codevar{end}}
                \State write out values $\mathbf{v_2}$[\codevar{start}] through $\mathbf{v_2}$[\codevar{end}]
            \EndIf
        \EndFunction
    \end{algorithmic}
\end{algorithm}

\subsubsection{\texorpdfstring{$\bm{L_{1,0}}$}{L} and its Recursion}
$L_{1, 0}$ calculates the third case ($\head{a}=0, \head{b}=1$) from $F_1$ (see (\ref{relwork:F1})). A similar recursion as for $L_{0, 1}$ can be defined for $L_{1, 0}$. However, when the second bit of $a$ is 1, the string pair will access values above $2^{2\ell-1}-1$. In this case, we use (\ref{impl:comp_eq}) to transform the value accesses to their symmetric position within the vector.

Essentially, these recursions serve as wrappers for the algorithm, dictating the portions of $\mathbf{v_2}$ to calculate and the values from $\mathbf{v_1}$ that must be loaded in to facilitate those calculations. By dividing the calculations into specific contiguous chunks, the recursions guarantee that disk memory is read from sequentially, with no unnecessary values read in, while maintaining the property that disk memory is written to in sequential chunks. Additional optimizations to these algorithms with substantial performance impacts are discussed in 
\cref{appx1}.
\section{Results} \label{sec:results}
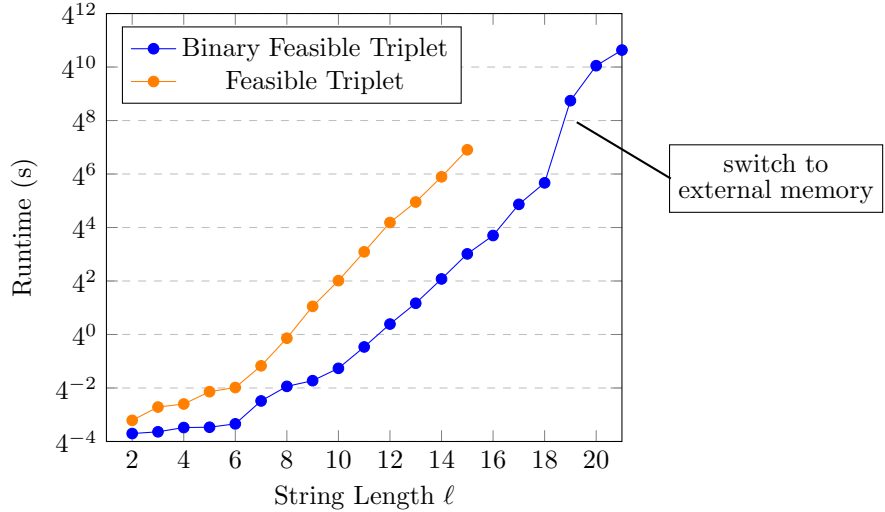
\begin{figure*}[t]
    \centering
    \begin{tikzpicture}
        \begin{axis}[
            legend pos=north west,
            xlabel={String Length $\ell$},
            ylabel={Runtime (s)},
            xtick distance = 2,
            ytick distance = 4^2,
            xmin=1, xmax=21,
            ymin=4^-4, ymax=4^12,
            ymajorgrids=true,
            grid style=dashed,
            /pgf/number format/.cd,
            1000 sep={},
            ymode=log,
            log basis y={4} 
        ]

            \addplot[color=blue, mark=*,]
                coordinates {
                (2,0.00589784)(3,0.00645251)(4,0.00805009)(5,0.00818167)(6,0.00974104)(7,0.03192200)(8,0.06803020)(9,0.09131590)(10,0.173263)(11,0.524698)(12,1.71853)(13,5.055211)(14,17.8179)(15,65.2633)(16,169.29)(17,850.775786)(18,2592.195)(19,183431)(20,1124200.26016)
                (21,2528289.71433)
                };
            \addplot[color=orange, mark=*,]
                coordinates {
                    (2,0.0116822)(3,0.0232844)(4,0.0272557)(5,0.0516464)(6,0.0640869)(7,0.197367)(8,0.825206)(9,4.30357)(10,16.2998)(11,72.604)(12,332.064)(13,960.543)(14,3546.34)(15,14390.7)
                };

            \legend{Binary Feasible Triplet, Feasible Triplet}
        \end{axis}
        
        \draw[black, thick] (7.5,3.5) -- (6.25,4.25);
        \node[draw, align=center, font=\fontsize{10pt}{10pt}\selectfont] at (8.9,3.5) {switch to \\ external memory};
    \end{tikzpicture}
    \caption{Runtime of the Feasible Triplet and Binary Feasible Triplet algorithms for $\gamma_{2,2}$ as string length parameter $\ell$ increases. Specialization to the binary case improves performance and the presented method of memory I/O reduces the overhead incurred by switching from RAM to disk memory to less than 18x.}
    \label{fig:runtime}
\end{figure*}

In our experiments we ran \cref{alg:feasibletriplet} for various values of $\sigma$, $d$, and $\ell$, and \cref{alg:bft} for increasingly large values of $\ell$. The results from our binary-case computations compared to the results obtained by Lueker \cite{Lueker2009} are shown in \cref{tab:binaryresults}. The amounts of time it took to achieve these bounds are displayed in \cref{fig:runtime}. When reported in the tables, lower bounds are rounded down to their 6th decimal place. To our knowledge, for $\ell\ge16$, these bounds exceed any previously reported lower bounds on $\gamma$. We also believe that for all but $\sigma=14, d=2$, the results in \cref{tab:generalresults} exceed all previously reported lower bounds for $\gamma_{\sigma, d}$.

\begin{table}[H]
    \centering
    \caption{Lower bounds on $\gamma$ for values of $\ell$, the string length parameter, from 1 to 21. Bold numbers represent new best lower bounds. \\}
    \begin{tabular}{lcc}
        \hline\hline
        \textbf{$\ell$} & \multicolumn{1}{c}{\textbf{\begin{tabular}[c]{@{}c@{}}Lower bound on $\gamma$ \\ (Lueker \cite{Lueker2009})\end{tabular}}} & \multicolumn{1}{c}{\textbf{\begin{tabular}[c]{@{}c@{}}Lower bound on $\gamma$\\ (ours)\end{tabular}}}  \\ \hline
        1 & 0.666666 & 0.666666 \\
        2 & 0.727272 & 0.727272 \\
        3 & 0.747922 & 0.747922 \\
        4 & 0.758576 & 0.758576 \\
        5 & 0.765446 & 0.765446 \\
        6 & 0.770273 & 0.770273  \\
        7 & 0.773975 & 0.773975 \\
        8 & 0.776860 & 0.776860 \\
        9 & 0.779259 & 0.779259 \\
        10 & 0.781281 & 0.781281 \\
        11 & 0.783005 & 0.783005 \\
        12 & 0.784515 & 0.784515 \\
        13 & 0.785841 & 0.785841 \\
        14 & 0.787017 & 0.787017 \\
        15 & 0.788071 & 0.788071 \\
        16 & - & \textbf{0.789021} \\
        17 & - & \textbf{0.789882} \\
        18 & - & \textbf{0.790668} \\
        19 & - & \textbf{0.791389} \\
        20 & - & \textbf{0.792052} \\ 
        21 & - & \textbf{0.792665} \\
        \hline\hline
    \end{tabular}
    \label{tab:binaryresults}
\end{table}

\begin{table*}
    \centering
    \caption{Lower bounds for $\gamma_{\sigma, d}$ with the value of the string length parameter $\ell$ we set to achieve our bounds. Bold numbers represent new best lower bounds.}
    \vspace*{1em}
    \begin{minipage}[t]{0.475\linewidth}
        \vspace*{-\topskip}
        \centering
\begin{tabular}{lccc}
\hline \hline
\multicolumn{4}{c}{Alphabet size $\sigma = 2$}       \\ \hline
                    & \multicolumn{2}{c}{Lower bound on $\gamma_{2,d}$} &                                    \\
{$d$} & Previous best              & Our results              & $\ell$ \\ \hline
2  & 0.788071 & \textbf{0.792665}               & 21 \\
3  & 0.704473 & \textbf{0.711548}               & 10 \\
4  & 0.661274 & \textbf{0.664722}               & 6  \\
5  & 0.636022 & \textbf{0.639248}               & 5  \\
6  & 0.617761 & \textbf{0.621057}               & 4  \\
7  & 0.602493 & \textbf{0.607261}               & 3  \\
8  & 0.594016 & \textbf{0.598782}               & 3  \\
9  & 0.587900 & \textbf{0.592177}               & 3  \\
10 & 0.570155 & \textbf{0.582348}               & 2  \\
11 & 0.570155 & \textbf{0.578463}               & 2  \\
12 & 0.563566 & \textbf{0.574268}               & 2  \\
13 & 0.563566 & \textbf{0.571067} & 2  \\
14 & 0.558494 & 0.558494                        & 1  \\
15 & -        & \textbf{0.558494}               & 1 \\
\hline \hline
\multicolumn{4}{c}{Alphabet size $\sigma = 3$} \\ \hline
                    & \multicolumn{2}{c}{Lower bound on $\gamma_{3,d}$}    &                                    \\
$d$ & Previous best & Our results & $\ell$ \\ \hline
2    & 0.671697   & \textbf{0.682218}   & 9  \\
3    & 0.556649   & \textbf{0.564841}   & 5  \\
4    & 0.498525   & \textbf{0.509237}   & 4  \\
5    & 0.461402   & \textbf{0.474304}   & 3  \\
6    & 0.421436   & \textbf{0.445434}   & 2  \\
7    & 0.413611   & \textbf{0.434513}   & 2  \\
8    & 0.405539   & \textbf{0.425774}   & 2 \\
9    & -          & \textbf{0.400949}   & 1 \\
\hline \hline
\multicolumn{4}{c}{Alphabet size $\sigma = 4$}               \\ \hline
 & \multicolumn{2}{c}{Lower bound on $\gamma_{4,d}$} & \\
$d$ & Previous best & Our results       &  $\ell$ \\ \hline
2                    & 0.599248      & \textbf{0.614333} & 8 \\
3                    & 0.457311      & \textbf{0.472979} & 4 \\
4                    & 0.389008      & \textbf{0.405702} & 3 \\
5                    & 0.335517      & \textbf{0.365329} & 2 \\
6                    & 0.324014      & \textbf{0.349848} & 2 \\
7                    & -             & \textbf{0.317032} & 1 \\
\hline \hline
\end{tabular}
\end{minipage}%
\begin{minipage}[t]{0.05\linewidth}
    \hspace*{0.05\linewidth}
\end{minipage}%
\begin{minipage}[t]{0.475\linewidth}
    \vspace*{-\topskip}
    \centering
\begin{tabular}{lccc}
\hline \hline
\multicolumn{4}{c}{Alphabet size $\sigma = 5$}               \\ \hline
  & \multicolumn{2}{c}{Lower bound on $\gamma_{5,d}$} &   \\
$d$ & Previous best & Our results       & $\ell$  \\ \hline
2                    & 0.539129      & \textbf{0.549817} & 5 \\
3                    & 0.356717      & \textbf{0.394945} & 3 \\
4                    & 0.289398      & \textbf{0.324337} & 2 \\
5                    & 0.273884      & \textbf{0.302235} & 2 \\
6                    & -             & \textbf{0.263369} & 1 \\
\hline \hline
\multicolumn{4}{c}{Alphabet size $\sigma = 6$}               \\ \hline
 & \multicolumn{2}{c}{Lower bound on $\gamma_{6,d}$} &   \\
$d$ & Previous best & Our results       &  $\ell$ \\ \hline
2                    & 0.479452      & \textbf{0.499229} & 4 \\
3                    & 0.309424      & \textbf{0.347798} & 3 \\
4                    & 0.245283      & \textbf{0.277835} & 2 \\
5                    & -             & \textbf{0.231234} & 1 \\
\hline \hline
\multicolumn{4}{c}{Alphabet size $\sigma = 7$}               \\ \hline
  & \multicolumn{2}{c}{Lower bound on $\gamma_{7,d}$} &  \\
$d$ & Previous best & Our results       & $\ell$  \\ \hline
2                    & 0.44502       & \textbf{0.466481} & 4 \\
3                    & 0.234567      & \textbf{0.273275} & 2 \\
4                    & 0.212786      & \textbf{0.242798} & 2 \\
5                    & -             & \textbf{0.200004} & 1 \\
\hline \hline
\multicolumn{4}{c}{Alphabet size $\sigma = 8$}               \\ \hline
 & \multicolumn{2}{c}{Lower bound on $\gamma_{8,d}$} & \\
$s$ & Previous best & Our results       &  $\ell$ \\ \hline
2                    & 0.42237       & \textbf{0.438799} & 4 \\
3                    & 0.207547      & \textbf{0.244709} & 2 \\
4                    & -             & \textbf{0.187869} & 1 \\
\hline \hline
\multicolumn{4}{c}{Alphabet size $\sigma = 9$}               \\ \hline
 & \multicolumn{2}{c}{Lower bound on $\gamma_{9,d}$} &  \\
$d$ & Previous best & Our results       &  $\ell$ \\ \hline
2                    & 0.40321       & \textbf{0.414876} & 4 \\
3                    & 0.186104      & \textbf{0.221554} & 2 \\
4                    & -             & \textbf{0.168164} & 1 \\
\hline \hline
\multicolumn{4}{c}{Alphabet size $\sigma = 10$}              \\ \hline
 & \multicolumn{2}{c}{Lower bound on $\gamma_{10,d}$} &  \\
$d$ & Previous best & Our results       &  $\ell$ \\ \hline
2                    & 0.38656       & \textbf{0.393811} & 4 \\
3                    & 0.168674      & \textbf{0.202401} & 2 \\
4                    & -             & \textbf{0.152193} & 1 \\
\hline \hline
\end{tabular}
\end{minipage}%
\label{tab:generalresults}
\end{table*}

The initial code to generate these bounds was written completely independently from the codebase linked in \cite{Lueker2009}. We also developed a separate code base containing a C++ implementation of Lueker's algorithm converted directly from his code (see \cref{appx2}). The bounds from all three independently developed code bases match exactly for $\ell \le 15$, indicating that the implementations are likely correct. Our results from the general case also exactly match Kiwi and Soto's \cite{KiwiSoto2009} for matching values of $\sigma$, $d$, and $\ell$ (see \cref{appx3}).
\newpage
\section{Conclusion}\label{sec:conc}
By augmenting previous algorithms for computing lower bounds with run-time optimizations, parallelization, and a recursive memory I/O scheme to allow sequential reading and writing once values become to large to store in RAM, we established new best lower bounds on the Chv\'{a}tal-Sankoff constants in the classical two binary string case as well as in the extended, general cases. All code is open source and available at \url{https://github.com/Statistics-of-Subsequences/PaperMaterials}.

While approximations of the value of the Chv\'{a}tal-Sankoff constant for pairs of binary strings have been made (see \cref{tab:bounds-estimates}), it seems that most of the work for the general case has been focused on tightening the upper and lower bounds. We believe that performing calculations to obtain precise approximations of different $\gamma_{\sigma, d}$ could be an avenue for future research yielding noteworthy results. Kiwi and Soto, as well as Steele, express interest in finding a relationship between different $\gamma_{\sigma, d}$ \cite{KiwiSoto2009, Steele1986}. We agree that such a relationship, if it exists, would be interesting, and also wonder if having precise approximations for the constants would aid in finding a relationship. We also believe that optimizing algorithms which generate upper bounds on the Chv\'{a}tal-Sankoff constants may be an attractive direction for future work.

\newpage \vspace*{-3em}
\bibliography{references.bib}

\newpage \appendix
\begin{appendices}
    \section{\small \raggedright{Program Code and Additional Material}}\label{appx1}
All program code is open source and available at \url{https://github.com/Statistics-of-Subsequences/PaperMaterials}. This link also contains instructions on how to run the code.

The files hosted at this domain also include explanations of additional optimizations we implemented which were too lengthy to include in this paper. These optimizations focus on the Binary Feasible Triplet Algorithm (\cref{alg:bft}), though we also note how it is possible to efficiently `pre-compute' the $\mathbf{b}$ vector in the general case. In particular, by intelligently combining $L_{0,1}$ and $L_{1,0}$, computing $L_{0,0}$ in a slightly less straightforward manner, reusing partial computations, and performing \texttt{max} operations locally wherever possible, these optimizations significantly reduce either A: the amount of memory I/O required each iteration, or B: the amount of computation performed each iteration, depending on the implementation.

\section{\small \raggedright{Alternate Lower Bound Algorithm}} \label{appx2}
We also developed a separate algorithm based on the implementation Lueker provided (see \cite{Lueker2009} for Lueker's implementation). This algorithm was a preliminary implementation of our modified Binary Feasible Triplet algorithm and obtained the same results up to $\ell = 18$, supporting the results given in \cref{tab:binaryresults}.

\section{\small \raggedright{Complete Results for General $\gamma_{\sigma, d}$ Constants}}\label{appx3}
Here we list out all of our results for the Feasible Triplet algorithm for all $\sigma$, $d$ values, even for smaller values of $\ell$ which did not result in better lower bounds. These results can be cross-checked with Table 2 of \cite{KiwiSoto2009} to demonstrate that our implementation of \cref{alg:feasibletriplet} results in the exact same numbers for the values of $\ell$ reported by Kiwi and Soto.

\begin{table*}[!ht]
    \centering
        \begin{tabular}{|C{0.10125in}C{0.17375in}C{0.525in}C{0.525in}C{0.525in}C{0.525in}C{0.525in}C{0.525in}C{0.525in}C{0.525in}C{0.525in}|}
            \hline
            \rowcolor{tableHeader2} \multicolumn{2}{|c}{\cellcolor{tableHeader1}{\color[HTML]{FFFFFF} }}                           & \multicolumn{9}{c|}{\cellcolor{tableHeader2} $\sigma$}                                           \\
            \rowcolor{tableHeader3} \multicolumn{2}{|c}{\multirow{-2}{*}{\cellcolor{tableHeader1}{\color[HTML]{FFFFFF}$\ell=1$}}}  & 2        & 3        & 4        & 5        & 6        & 7        & 8        & 9        & 10       \\
            \cellcolor{tableHeader2}                       & \cellcolor{tableHeader3}2                                             & 0.666666 & 0.500000 & 0.400000 & 0.333333 & 0.285714 & 0.250000 & 0.222222 & 0.200000 & 0.181818 \\
            \cellcolor{tableHeader2}                       & \cellcolor{tableHeader3}3                                             & 0.666666 & 0.488372 & 0.384615 & 0.317073 & 0.269662 & 0.234567 & 0.207547 & 0.186104 & 0.168674 \\
            \cellcolor{tableHeader2}                       & \cellcolor{tableHeader3}4                                             & 0.666666 & 0.450000 & 0.352583 & 0.289398 & 0.245283 & 0.212786 & 0.187869 & 0.168164 & 0.152193 \\
            \cellcolor{tableHeader2}                       & \cellcolor{tableHeader3}5                                             & 0.666666 & 0.432494 & 0.335517 & 0.273884 & 0.231234 & 0.200004 &          &          &          \\
            \cellcolor{tableHeader2}                       & \cellcolor{tableHeader3}6                                             & 0.592592 & 0.421434 & 0.324014 & 0.263369 &          &          &          &          &          \\
            \cellcolor{tableHeader2}                       & \cellcolor{tableHeader3}7                                             & 0.592592 & 0.413611 & 0.317032 &          &          &          &          &          &          \\
            \cellcolor{tableHeader2}                       & \cellcolor{tableHeader3}8                                             & 0.579185 & 0.405539 &          &          &          &          &          &          &          \\
            \cellcolor{tableHeader2}                       & \cellcolor{tableHeader3}9                                             & 0.579185 & 0.400949 &          &          &          &          &          &          &          \\
            \cellcolor{tableHeader2}                       & \cellcolor{tableHeader3}10                                            & 0.570155 &          &          &          &          &          &          &          &          \\
            \cellcolor{tableHeader2}                       & \cellcolor{tableHeader3}11                                            & 0.570155 &          &          &          &          &          &          &          &          \\
            \cellcolor{tableHeader2}                       & \cellcolor{tableHeader3}12                                            & 0.563566 &          &          &          &          &          &          &          &          \\
            \cellcolor{tableHeader2}                       & \cellcolor{tableHeader3}13                                            & 0.563566 &          &          &          &          &          &          &          &          \\
            \cellcolor{tableHeader2}                       & \cellcolor{tableHeader3}14                                            & 0.558494 &          &          &          &          &          &          &          &          \\
            \multirow{-14}{*}{\cellcolor{tableHeader2}$d$} & \cellcolor{tableHeader3}15                                            & 0.558494 &          &          &          &          &          &          &          &          \\
            \hline
        \end{tabular}
\end{table*}

\begin{table*}[!ht]
    \centering
        \begin{tabular}{|C{0.10125in}C{0.17375in}C{0.525in}C{0.525in}C{0.525in}C{0.525in}C{0.525in}C{0.525in}C{0.525in}C{0.525in}C{0.525in}|}
            \hline
            \rowcolor{tableHeader2} \multicolumn{2}{|c}{\cellcolor{tableHeader1}{\color[HTML]{FFFFFF} }}                           & \multicolumn{9}{c|}{\cellcolor{tableHeader2} $\sigma$}                                           \\
            \rowcolor{tableHeader3} \multicolumn{2}{|c}{\multirow{-2}{*}{\cellcolor{tableHeader1}{\color[HTML]{FFFFFF}$\ell=2$}}}  & 2        & 3        & 4        & 5        & 6        & 7        & 8        & 9        & 10       \\
            \cellcolor{tableHeader2}                       & \cellcolor{tableHeader3}2                                             & 0.727273 & 0.620690 & 0.542373 & 0.480769 & 0.431138 & 0.390438 & 0.356545 & 0.327935 & 0.303490 \\
            \cellcolor{tableHeader2}                       & \cellcolor{tableHeader3}3                                             & 0.673913 & 0.516896 & 0.421518 & 0.356717 & 0.309424 & 0.273275 & 0.244710 & 0.221555 & 0.202402 \\
            \cellcolor{tableHeader2}                       & \cellcolor{tableHeader3}4                                             & 0.643216 & 0.484937 & 0.389008 & 0.324338 & 0.277835 & 0.242798 &          &          &          \\
            \cellcolor{tableHeader2}                       & \cellcolor{tableHeader3}5                                             & 0.626506 & 0.461402 & 0.365329 & 0.302236 &          &          &          &          &          \\
            \cellcolor{tableHeader2}                       & \cellcolor{tableHeader3}6                                             & 0.610925 & 0.445434 & 0.349848 &          &          &          &          &          &          \\
            \cellcolor{tableHeader2}                       & \cellcolor{tableHeader3}7                                             & 0.602493 & 0.434514 &          &          &          &          &          &          &          \\
            \cellcolor{tableHeader2}                       & \cellcolor{tableHeader3}8                                             & 0.594016 & 0.425774 &          &          &          &          &          &          &          \\
            \cellcolor{tableHeader2}                       & \cellcolor{tableHeader3}9                                             & 0.587900 &          &          &          &          &          &          &          &          \\
            \cellcolor{tableHeader2}                       & \cellcolor{tableHeader3}10                                            & 0.582349 &          &          &          &          &          &          &          &          \\
            \cellcolor{tableHeader2}                       & \cellcolor{tableHeader3}11                                            & 0.578464 &          &          &          &          &          &          &          &          \\
            \cellcolor{tableHeader2}                       & \cellcolor{tableHeader3}12                                            & 0.574269 &          &          &          &          &          &          &          &          \\
            \multirow{-12}{*}{\cellcolor{tableHeader2}$d$} & \cellcolor{tableHeader3}13                                            & 0.571067 &          &          &          &          &          &          &          &          \\
            \hline
        \end{tabular}
\end{table*}

\begin{table*}[!ht]
    \centering
        \begin{tabular}{|C{0.10125in}C{0.17375in}C{0.525in}C{0.525in}C{0.525in}C{0.525in}C{0.525in}C{0.525in}C{0.525in}C{0.525in}C{0.525in}|}
            \hline
            \rowcolor{tableHeader2} \multicolumn{2}{|c}{\cellcolor{tableHeader1}{\color[HTML]{FFFFFF} }}                          & \multicolumn{9}{c|}{\cellcolor{tableHeader2} $\sigma$}                                           \\
            \rowcolor{tableHeader3} \multicolumn{2}{|c}{\multirow{-2}{*}{\cellcolor{tableHeader1}{\color[HTML]{FFFFFF}$\ell=3$}}} & 2        & 3        & 4        & 5        & 6        & 7        & 8        & 9        & 10       \\
            \cellcolor{tableHeader2}                      & \cellcolor{tableHeader3}2                                             & 0.747922 & 0.644966 & 0.573254 & 0.521091 & 0.479452 & 0.444577 & 0.414651 & 0.388537 & 0.365485 \\
            \cellcolor{tableHeader2}                      & \cellcolor{tableHeader3}3                                             & 0.687410 & 0.545373 & 0.457311 & 0.394945 & 0.347798 &          &          &          &          \\
            \cellcolor{tableHeader2}                      & \cellcolor{tableHeader3}4                                             & 0.651309 & 0.498525 & 0.405702 &          &          &          &          &          &          \\
            \cellcolor{tableHeader2}                      & \cellcolor{tableHeader3}5                                             & 0.632165 & 0.474304 &          &          &          &          &          &          &          \\
            \cellcolor{tableHeader2}                      & \cellcolor{tableHeader3}6                                             & 0.617761 &          &          &          &          &          &          &          &          \\
            \cellcolor{tableHeader2}                      & \cellcolor{tableHeader3}7                                             & 0.607261 &          &          &          &          &          &          &          &          \\
            \cellcolor{tableHeader2}                      & \cellcolor{tableHeader3}8                                             & 0.598782 &          &          &          &          &          &          &          &          \\
            \multirow{-8}{*}{\cellcolor{tableHeader2}$d$} & \cellcolor{tableHeader3}9                                             & 0.592177 &          &          &          &          &          &          &          &          \\
            \hline
        \end{tabular}
\end{table*}

\begin{table*}[!ht]
    \centering
        \begin{tabular}{|C{0.10125in}C{0.17375in}C{0.525in}C{0.525in}C{0.525in}C{0.525in}C{0.525in}C{0.525in}C{0.525in}C{0.525in}C{0.525in}|}
            \hline
            \rowcolor{tableHeader2} \multicolumn{2}{|c}{\cellcolor{tableHeader1}{\color[HTML]{FFFFFF} }}                          & \multicolumn{9}{c|}{\cellcolor{tableHeader2} $\sigma$}                                           \\
            \rowcolor{tableHeader3} \multicolumn{2}{|c}{\multirow{-2}{*}{\cellcolor{tableHeader1}{\color[HTML]{FFFFFF}$\ell=4$}}} & 2        & 3        & 4        & 5        & 6        & 7        & 8        & 9        & 10       \\
            \cellcolor{tableHeader2}                      & \cellcolor{tableHeader3}2                                             & 0.758576 & 0.657642 & 0.589484 & 0.539129 & 0.499229 & 0.466481 & 0.438799 & 0.414876 & 0.393811 \\
            \cellcolor{tableHeader2}                      & \cellcolor{tableHeader3}3                                             & 0.692950 & 0.556649 & 0.472979 &          &          &          &          &          &          \\
            \cellcolor{tableHeader2}                      & \cellcolor{tableHeader3}4                                             & 0.657241 & 0.509237 &          &          &          &          &          &          &          \\
            \cellcolor{tableHeader2}                      & \cellcolor{tableHeader3}5                                             & 0.636022 &          &          &          &          &          &          &          &          \\
            \multirow{-5}{*}{\cellcolor{tableHeader2}$d$} & \cellcolor{tableHeader3}6                                             & 0.621057 &          &          &          &          &          &          &          &          \\
            \hline
        \end{tabular}
\end{table*}

\begin{table*}[!ht]
    \centering
        \begin{tabular}[t]{|C{0.10125in}C{0.17375in}C{0.525in}C{0.525in}C{0.525in}C{0.525in}|}
            \hline
            \rowcolor{tableHeader2} \multicolumn{2}{|c}{\cellcolor{tableHeader1}{\color[HTML]{FFFFFF} }}                          & \multicolumn{4}{c|}{\cellcolor{tableHeader2} $\sigma$} \\
            \rowcolor{tableHeader3} \multicolumn{2}{|c}{\multirow{-2}{*}{\cellcolor{tableHeader1}{\color[HTML]{FFFFFF}$\ell=5$}}} & 2        & 3        & 4        & 5                     \\
            \cellcolor{tableHeader2}                      & \cellcolor{tableHeader3}2                                             & 0.765446 & 0.665874 & 0.599248 & 0.549817              \\
            \cellcolor{tableHeader2}                      & \cellcolor{tableHeader3}3                                             & 0.697737 & 0.564841 &          &                       \\
            \cellcolor{tableHeader2}                      & \cellcolor{tableHeader3}4                                             & 0.661274 &          &          &                       \\
            \multirow{-4}{*}{\cellcolor{tableHeader2}$d$} & \cellcolor{tableHeader3}5                                             & 0.639248 &          &          &                       \\
            \hline
        \end{tabular}
    \quad
        \begin{tabular}[t]{|C{0.10125in}C{0.17375in}C{0.525in}C{0.525in}C{0.525in}|}
            \hline
            \rowcolor{tableHeader2} \multicolumn{2}{|c}{\cellcolor{tableHeader1}{\color[HTML]{FFFFFF} }}                          & \multicolumn{3}{c|}{\cellcolor{tableHeader2} $\sigma$} \\
            \rowcolor{tableHeader3} \multicolumn{2}{|c}{\multirow{-2}{*}{\cellcolor{tableHeader1}{\color[HTML]{FFFFFF}$\ell=6$}}} & 2        & 3        & 4                                \\
            \cellcolor{tableHeader2}                      & \cellcolor{tableHeader3}2                                             & 0.770273 & 0.671697 & 0.605786                         \\
            \cellcolor{tableHeader2}                      & \cellcolor{tableHeader3}3                                             & 0.701317 &          &                                  \\
            \multirow{-3}{*}{\cellcolor{tableHeader2}$d$} & \cellcolor{tableHeader3}4                                             & 0.664722 &          &                                  \\
            \hline
        \end{tabular}
\end{table*}

\begin{table*}[!ht]
    \centering
        \begin{tabular}[t]{|C{0.10125in}C{0.17375in}C{0.525in}C{0.525in}C{0.525in}|}
            \hline
            \rowcolor{tableHeader2} \multicolumn{2}{|c}{\cellcolor{tableHeader1}{\color[HTML]{FFFFFF} }}                          & \multicolumn{3}{c|}{\cellcolor{tableHeader2} $\sigma$} \\
            \rowcolor{tableHeader3} \multicolumn{2}{|c}{\multirow{-2}{*}{\cellcolor{tableHeader1}{\color[HTML]{FFFFFF}$\ell=7$}}} & 2        & 3        & 4                                \\
            \cellcolor{tableHeader2}                      & \cellcolor{tableHeader3}2                                             & 0.773975 & 0.676041 & 0.610590                         \\
            \multirow{-2}{*}{\cellcolor{tableHeader2}$d$} & \cellcolor{tableHeader3}3                                             & 0.704473 &          &                                  \\
            \hline
        \end{tabular}
    \quad
        \begin{tabular}[t]{|C{0.10125in}C{0.17375in}C{0.525in}C{0.525in}C{0.525in}|}
            \hline
            \rowcolor{tableHeader2} \multicolumn{2}{|c}{\cellcolor{tableHeader1}{\color[HTML]{FFFFFF} }}                          & \multicolumn{3}{c|}{\cellcolor{tableHeader2} $\sigma$} \\
            \rowcolor{tableHeader3} \multicolumn{2}{|c}{\multirow{-2}{*}{\cellcolor{tableHeader1}{\color[HTML]{FFFFFF}$\ell=8$}}} & 2        & 3        & 4                                \\
            \cellcolor{tableHeader2}                      & \cellcolor{tableHeader3}2                                             & 0.776860 & 0.679441 & 0.614333                         \\
            \multirow{-2}{*}{\cellcolor{tableHeader2}$d$} & \cellcolor{tableHeader3}3                                             & 0.707165 &          &                                  \\
            \hline
        \end{tabular}
\end{table*}

\begin{table*}[!ht]
    \centering
        \begin{tabular}{|C{0.10125in}C{0.17375in}C{0.525in}C{0.525in}|}
            \hline
            \rowcolor{tableHeader2} \multicolumn{2}{|c}{\cellcolor{tableHeader1}{\color[HTML]{FFFFFF} }}                          & \multicolumn{2}{c|}{\cellcolor{tableHeader2} $\sigma$} \\
            \rowcolor{tableHeader3} \multicolumn{2}{|c}{\multirow{-2}{*}{\cellcolor{tableHeader1}{\color[HTML]{FFFFFF}$\ell=9$}}} & 2        & 3                                           \\
            \cellcolor{tableHeader2}                      & \cellcolor{tableHeader3}2                                             & 0.779259 & 0.682218                                    \\
            \multirow{-2}{*}{\cellcolor{tableHeader2}$d$} & \cellcolor{tableHeader3}3                                             & 0.709501 &                                             \\
            \hline
        \end{tabular}
    \quad
        \begin{tabular}{|C{0.10125in}C{0.17375in}C{0.525in}|}
            \hline
            \rowcolor{tableHeader2} \multicolumn{2}{|c}{\cellcolor{tableHeader1}{\color[HTML]{FFFFFF} }}                           & $\sigma$ \\
            \rowcolor{tableHeader3} \multicolumn{2}{|c}{\multirow{-2}{*}{\cellcolor{tableHeader1}{\color[HTML]{FFFFFF}$\ell=10$}}} & 2        \\
            \cellcolor{tableHeader2}                      & \cellcolor{tableHeader3}2                                              & 0.781281 \\
            \multirow{-2}{*}{\cellcolor{tableHeader2}$d$} & \cellcolor{tableHeader3}3                                              & 0.711548 \\
            \hline
        \end{tabular}
\end{table*}

\newpage \vspace*{100em} \textcolor{white}{This text is only here so that the tables will not be spread out across the entire height of the page}
\end{appendices}

\end{document}